\documentstyle[aps,pra,epsf]{revtex}

\def\be{\begin{equation}}
\def\bea{\begin{eqnarray}}
\def\bma{\begin{mathletters}}
\def\ee{\end{equation}}
\def\eea{\end{eqnarray}}
\def\ema{\end{mathletters}}

\title{Mean Field Approximations and Multipartite Thermal Correlations}
\author{Vlatko Vedral}
\address{Optics Section, Blackett Laboratory, Imperial College,\\
Prince Consort Road SW7 2BZ}
\date{\today}

\begin{document}

\maketitle

\begin{abstract}
The relationship between the mean-field approximations in various interacting models
of statistical physics and measures of classical and quantum correlations is explored.
We present a method that allows us to bound the total amount of correlations (and
hence entanglement) in a physical system in thermal equilibrium at some temperature in
terms of its free energy and internal energy. This method is first illustrated using
two qubits interacting through the Heisenberg coupling, where entanglement and
correlations can be computed exactly. It is then applied to the one dimensional Ising
model in a transverse magnetic field, for which entanglement and correlations cannot
be obtained by exact methods. We analyze the behavior of correlations in various
regimes and identify critical regions, comparing them with already known results.
Finally, we present a general discussion of the effects of entanglement on the
macroscopic, thermodynamical features of solid-state systems. In particular, we
exploit the fact that a $d$ dimensional quantum system in thermal equilibrium can be
made to corresponds to a $d+1$ classical system in equilibrium to substitute all
entanglement for classical correlations.
\end{abstract}

\section{Introduction}

Entanglement is an effect that has been under a great deal of scrutiny since $1935$,
when Einstein, Podolsky and Rosen and (independently) Schr\"odinger introduced it as a
purely quantum phenomenon without any counterpart in classical physics. Although
counterintuitive by its nature, entanglement is a real physical effect that has been
experimentally demonstrated through violation of Bell's inequalities, experiments on
teleportation, dense coding and entanglement swapping. Most recently, entanglement has
been shown to affect macroscopic properties of solids, such as its magnetization and
heat capacity \cite{Nature}. This effect is observed only at low temperature,
nevertheless, it is still impressive that a virtually macroscopic system can feel
effects of purely quantum correlations. The purpose of this paper is to explore a
general method of quantifying and bounding total correlations in any such physical
system.

A great deal of effort has gone into theoretically understanding and quantifying
entanglement \cite{Vedral}. There is a large number of different measures proposed,
however, we have recently argued that based on the very natural local processes and
classical communication we should not expect to find a unique measure of entanglement
even for mixed bi-partite states \cite{Fumiaki}. Therefore, different measures capture
different aspects of entanglement, but no measure can capture all aspects. This
apparent ambiguity will not present a problem for us here as will be seen shortly.

The main purpose of this paper is to study multipartite thermal entanglement in
general solid state systems. Many investigations have recently been conducted in this
direction \cite{NiTe,W00,OW01,ved01,ved02,WFS01,WaZa02,Wa02,rest,Os02,OsNi02,vid02},
and entanglement is found to be present under some circumstance and has been linked to
the existence of critical phenomena. However, all of the work so far uses some form or
other of bipartite measures of entanglement. This is because we have an almost
complete understanding of bipartite entanglement and can even compute its amount
exactly. In this paper, on the other hand, we will be addressing genuine multipartite
correlations. The measure that we will use is the relative entropy of entanglement
which is the only current measure that can be consistently used for any number of
particles and dimensions.

The key result of this paper will be a derivation of an upper bound on the amount of
total correlation in a thermal state of any macroscopic system. This will, of course,
be a bound on the amount of entanglement as well, since entanglement is a special case
(i.e. the quantum part) of correlations. We will apply the derived bound to two simple
interacting systems of qubits and analyse the limitations of our method.
Interestingly, in spite of the bound not being exact, we are still able to see regions
of criticality. At the end, we will discuss the correspondence between $d$ dimensional
quantum interaction systems and $d+1$ dimensional classical systems and examine its
implications on the existence of macroscopic thermal entanglement.

In the remaining part we introduce the measures of information
that will be used throughout. We first introduce the quantum (von
Neumann) mutual information, which refers to the correlation
between two quantum subsystems. The quantum mutual information
between the two subsystems $\rho_1$ and $\rho_2$ of the joint
state ${\rho}_{12}$ is defined as
\begin{eqnarray}
I (\rho_1:\rho_2 ;\rho_{12}) = S(\rho_1) + S(\rho_2) - S(\rho_{12})\;\; .
\end{eqnarray}
where $S(\rho) = - \mbox{tr} \rho \ln \rho$ is the von Neumann entropy. For us very
useful will be the fact that this quantity can be interpreted as a distance between
the two quantum states $\rho_1$ and $\rho_2$. We define the quantum relative entropy,
in a direct analogy with the classical (Shannon) relative entropy \cite{Vedral}. The
quantum relative entropy between the two states $\sigma$ and $\rho$ is defined as
\begin{eqnarray}
S (\sigma ||\rho) = \mbox{tr} \{\sigma (\ln \sigma - \ln \rho)\} \label{relent} \;\; .
\end{eqnarray}
This measure has a very important application in statistical discrimination of the two
states, but this role will not be of concern to us here (see \cite{Vedral}). What is
important is that the quantum mutual information can be understood as a distance of
the state $\rho_{12}$ to the uncorrelated state $\rho_1\otimes\rho_2$,
\begin{eqnarray}
I (\rho_1:\rho_2 ;\rho_{12}) = S (\rho_{12} || \rho_1\otimes\rho_2)\; . \nonumber
\end{eqnarray}
where $\mbox{tr}_1 \rho_{12} = \rho_2$ and $\mbox{tr}_2 \rho_{12} = \rho_1$ The larger
this value, the more correlated the state $\rho_{12}$. The main advantage of looking
at the mutual information as a distance measure, is that it can be generalized in this
way to many particles. Another advantage is that it can be applied to quantify
multipartite entanglement \cite{Vedral} as will be seen later. This applicability to
many parties is precisely what we need to be able to quantify correlations in a large
(macroscopic) piece of solid (in the thermodynamical limit, when the number of
constituents tends to infinity). A completely uncorrelated state of $n$ systems would
be of the form $\rho_1\otimes \rho_2\otimes ...\rho_n$, and the difference between a
given state $\rho$ and this state would represent the amount of correlation present.
So, the muliparty mutual information is given by \cite{Vedral}
\begin{equation}
I(\rho_1: \rho_2: ...\rho_n;\rho) = S(\rho ||\rho_1\otimes \rho_2\otimes ...\rho_n) =
\sum_i S(\rho_i) - S(\rho)
\end{equation}
Note that only the individual entropies and the total entropy are
involved in this calculation. There is no need to look at the two,
three or more subsystems and their entropies. Note also that this
presents an upper bound to total entanglement in the state $\rho$.
Intuitively speaking this is because entanglement is only a part
of total correlations, and so it has to be smaller than the total
correlations. Mathematically, we will show this in the text below.

We would like to discuss the role that these measures of information play in various
models of interacting systems in statistical physics. Before we do that, we first
review how entropies can be used to generalize and formalize the notion of the mean
field approximation.

\section{Mean field Approximation}

The term ``mean field approximation" can have various meanings in statistical
mechanics and it actually refers to a whole set of different approximations all with
different levels of accuracy (due to Weiss, Bethe, etc \cite{Kubo}). The main idea is
to ignore the fact that the constituents of the solid under investigation are
interacting and treat the interaction as an ``average effect". The difference between
approximations lies in the way we perform the averaging. In this section we explain
one method of doing so that will be used later in the paper. The aforementioned bound
presented here will, however, be valid for any mean field approximation. We will
discuss the limits and failures of this method in section VI, but see \cite{Kubo} for
a general discussion of the limits of any mean-filed approximation. We would like to
stress that we will not at all make any new contribution here to the mean field
approximation; our aim is to use the existing theory to quantify multipartite
correlations.

The key quantity in statistical mechanics that allows us to derive thermodynamical
properties of the system under study is free energy. Once this quantity is known we
can, by differentiation with respect to suitable parameters, derive all other
thermodynamical variables, such as pressure, energy, magnetization and so forth. The
free energy is defined as
\begin{equation}
F = -kT \ln Z
\end{equation}
where $k$ is the Boltzmann constant, $T$ is the temperature and $Z$ is the partition
function. In order to calculate the partition function we need to know the energy
levels of the system, $E_i$, and then perform the following sum-over-states:
\begin{equation}
Z = \sum_i e^{E_i/kT}\; .
\end{equation}
If the state we are considering is a quantum mechanical density matrix in a thermal
state, then the partition function is given by
\begin{equation}
Z = \mbox{tr} e^{-\beta H}
\end{equation}
where $H$ is the Hamiltonian and $\beta = 1/kT$. Only quantum models will concern us
here and so this is the expression of the partition function that will be used. The
partition function is easy to calculate if the energy levels are known, but for most
interesting cases this cannot be done exactly. Hence various approximations must be
used. Before going into the general theory of how this calculation of $Z$ is
performed, we first present a very simple example, to illustrate the main point.

Interacting systems are frequently very difficult to treat in statistical mechanics
simply because there are no known analytic methods for evaluating the partition
function in general. Therefore, we most likely have to resort to some kind of
approximations. One of the well known ways of doing this is to write the Hamiltonian
of the interacting system as a sum of non-interacting Hamiltonians for individual
spins. The effect of interaction is incorporated in as the average field produced by
all the other spins. Here is how this method works for a simple one dimensional Ising
model.

The Hamiltonian for this model is given by:
\begin{equation}
H = -\frac{J}{2} \sum_{i=1}^{N} \sigma_i^z\otimes \sigma_{i+1}^z - B \sum_{i=1}^{N}
\sigma_i^z \; .
\end{equation}
This model has been widely used to investigate various properties of real solids
\cite{Kubo} . It can be solved exactly, but for educational purposes we will
illustrate how to solve it using the mean field approximation. The mean field
Hamiltonian is
\begin{equation}
H_{MF} = -\frac{J}{2} \sum_{i=1}^{N} s \sigma_i^z - B
\sum_{i=1}^{N} \sigma_i^z
\end{equation}
where $s$ is the mean value of the neighboring spins and there are $2$ of them (but we
have also divided by $2$ not to double count!). This Hamiltonian is of course much
easier to treat than the original one and its eigenvalues can immediately be found to
be $\pm (sJ/2 + B)$. Therefore it is easy to obtain the mean field partition function
and hence all the other thermodynamical properties, which is why the mean field
approximation is so extensively used.

But, how do we find the value of $s$ which is necessary for being able to use this
approximation? The value of $s$ is given by $s=\mbox{tr}(\sigma^z H)$, but we have
assumed that $H$ is difficult to handle. So, instead, we use the following formula
\begin{equation}
s = \mbox{tr} (\sigma^z H_{MF})
\end{equation}
since $H_{MF}$ is very easy to handle. The dipole $s$ here really points in the $z$
direction only, but this is because we are considering a simple system. In the
subsequent sections we will have to compute averages of other Pauli operators as well.
This will result in a transcendental equation and for this problem in particular we
have that
\begin{equation}
s = \tanh \frac{(Js/2 + B)}{kT} \label{mean field}
\end{equation}
which cannot be solved analytically. When the external field is absent ($B=0$) we have
that the equality can be maintained only for some range of values of $J$. In fact,
this range is $J/2>kT$, which leads to a critical temperature of $T_c = J/2k$. Of
course, one dimensional models do not have phase transitions, and so the mean field
incorrectly predicts the existence of the critical temperature, but nevertheless, in
general this approximation turns out to be useful (especially in higher dimensions
\cite{Kubo}). We will use it to give us an upper bound on the amount of correlations
and entanglement present in interacting systems.

For convenience we will use the following dimensionless quantities in the paper $C=
B/kT$ and $K = J/2kT$. Note that this will imply, for example, that the limit of $C$
becoming large can be reached in two different ways: $B$ fixed and temperature tending
to zero, or, $T$ fixed and $B$ increasing to infinity. Although these are different
physically, they will invariably lead to the same result simply because only the ratio
of the field strength and temperature will appear in the formulae and never the two
quantities independently. We first discuss an important application of relative
entropy to mean field theory.

\section{Bogoliubov Re-derived}

We describe for the sake of completeness a general method for approximating the free
energy which is due to Bogoliubov. We will be interested not so much in his method for
computing the free energy as in the way that this can provide us with an upper bound
on total correlations in the system and hence its entanglement. Needless to say, this
was not Bogoliubov's original motivation.

We start from the fact that the relative entropy is always a positive quantity,
\begin{equation}
S(\rho ||\rho_{MF}) \ge 0
\end{equation}
where (in this paper) we specialise to thermal states of the form $\rho = e^{-\beta
H}/Z$ and $\rho_{MF} = e^{-\beta H_{MF}}/Z_{MF}$. The entropy of either of these
matrices is easy to compute. For example,
\begin{equation}
S(\rho) = -\mbox{tr} \frac{e^{-\beta H}}{Z} \ln \frac{e^{-\beta H}}{Z} = + \ln Z -
\beta \mbox{tr} (H \rho) \; .
\end{equation}
Note that the last term in the above equation is just $\beta$ times the average value
of energy $\langle H \rangle_H = \mbox{tr} (H\rho)$. The subscript $H$ in our notation
means that the average is done with respect to the state $\rho$ which is a thermal
state generated by the Hamiltonian $H$. It is now straightforward to see how to
compute the relative entropy between the real and mean field thermal states. After a
short calculation we have:
\begin{equation}
S(\rho ||\rho_{MF}) = \ln Z_{MF} - \ln Z + \beta \langle H_{MF} -
H \rangle_{H}
\end{equation}
where, as discussed before, $Z_{MF} = \mbox{tr} e^{\beta H_{MF}}$. From the positivity
of the relative entropy we therefore conclude that
\begin{equation}
0\le \ln Z_{MF} - \ln Z + \beta \langle H_{MF} - H \rangle_{H} \; .
\end{equation}
This is one of the Bogoliubov inequalities. For completeness, we also derive its
``mirror image". Namely, we can always use the relative entropy the other way round
$S(\rho_{MF} ||\rho) \ge 0$, which then implies that
\begin{equation}
0\le \ln Z - \ln Z_{MF} + \beta \langle H - H_{MF} \rangle_{H_{MF}} \; .
\end{equation}
We can combine the two inequalities into a single expression to
obtain
\begin{equation}
\beta \langle H - H_{MF} \rangle_{H} \le \ln Z_{MF} - \ln Z \le \beta \langle H -
H_{MF} \rangle_{H_{MF}} \; .
\end{equation}
We can also express this inequality via the free energy, $F= -\ln Z/\beta$, in which
case we obtain the so called Bogoliubov inequalities:
\begin{equation}
\beta \langle H_{MF} - H \rangle_{H_{MF}} \le F - F_{MF} \le \beta \langle H_{MF} - H
\rangle_{H} \; .
\end{equation}
The purpose of this is the following. Suppose that we would like to approximate the
partition function (or alternatively the free energy) because the true one cannot
easily be computed. This implies that ``free" parameters in the mean field Hamiltonian
should be varied so that the expression $F_{MF} + \beta \langle H_{MF} - H
\rangle_{H}$ is minimised. Since this expression has as its lower bound the true value
of the free energy (this is the statement of the Bogoliubov inequality), that means
that by finding the minimum we are performing the best possible approximation (within
this method, of course). We can alternatively maximize the left hand side. This method
is very general and has been used many times since exactly solvable models are very
rare in statistical mechanics. We will not directly be using this bound here, but will
instead show how to use it to bound the amount of correlations in any thermal state.

\section{Multipartite Relative Entropy of Entanglement}

We will now apply the formalism presented above to put a bound on the multipartite
mutual information and entanglement using the free energy and internal energy. This
will allow us to establish a relationship between the observable, macroscopic
thermodynamical quantities and the amount of correlations in a given solid.

As we have already stated, once we have the partition function all other
thermodynamical quantities can be derived \cite{Kubo}. For example, the magnetic
susceptibility is given by
\begin{equation}
\chi = \frac{\partial^2 \ln Z}{\partial B^2} \; .
\end{equation}
This quantity tells us the change in magnetization of the material as the external
field $B$ is increased. Microscopically, we have that the susceptibility is, in fact,
a sum over all microscopic spin correlation functions
\begin{equation}
\chi = \sum_{ij} \xi_{ij}
\end{equation}
where $\xi_{ij}$ is the spin correlation function between the sites $i$ and $j$. This
is a very important relation as it connects a macroscopic quantity $\chi$ to its
microscopic roots in the form of the two-site correlation functions $\xi_{ij}$. We
will show how entanglement, which fundamentally exists at the level of the microscopic
correlations function, can ``propagate" in its effects to the macroscopic level of
$\chi$. We will not be using the quantity $\xi_{ij}$ in any fundamental way in the
paper so that we will not need to give its precise definition. Suffice it to say that
any other thermodynamical quantity can in a similar way be derived from the Free
energy (partition function) and the interested reader is advised to consult further
any textbook on statistical mechanics such as \cite{Kubo}. Here we will only be
concerned will multiparty correlations whose link to thermodynamics is much clearer
than that of two-site entanglement.

We now define a particular measure of entanglement that has been extensively used in
the literature \cite{Vedral} and is also appropriate for our considerations. If ${\cal
D} $ is the set of all disentangled (separable) states (of the form $\sum_i p_i
\rho_1^1\otimes \rho_2^i$, where $p_i$ are some probabilities), the measure of
entanglement for a state $\sigma$ is then defined to be the Relative entropy of
entanglement
\begin{eqnarray}
E({\sigma}):= \min_{\rho \in {\cal D}}\,\,\, S(\sigma || \rho)
\label{6a}
\end{eqnarray}
where $S(\sigma || \rho)$ is the quantum relative entropy defined in eq.
(\ref{relent}). This measure tells us that the amount of entanglement present in the
state $\sigma$ is equal to its distance from the disentangled set of states. Note that
this  definition is, in spirit, a direct generalization of the mutual information.
Whereas mutual information measures the distance of the state to the closest
uncorrelated state, the relative entropy of entanglement measures its distance to the
closest classically (but not quantumly) correlated state. It is therefore immediately
clear that $E(\sigma) \le I(\sigma)$ (by definition).

We will now use the equation that we introduced in the previous section
\begin{equation}
S(\rho ||\rho_{MF}) = \ln Z_{MF} - \ln Z + \beta \langle H_{MF} -
H \rangle_{H}
\end{equation}
to show to what extent susceptibility is dependent on entanglement. Our consideration
is completely general and can be applied in any situation in statistical mechanics.
Suppose that the mean field theory gives us the closest separable state $\rho_{MF}$.
This is not a very good assumption in general as the state in mean field theory will
be completely uncorrelated and a separable state can still have some classical
correlations. This, in fact, is the advantage of the mean field approach: by
neglecting correlations we can then write down a Hamiltonian whose spectrum is much
easier to find (it boils down to diagonalising a two by two matrix for the one
dimensional Ising and Heisenberg models). However, if for the sake of argument, the
mean field gives us the closest separable state, then from the above equality we can
conclude that
\begin{equation}
E(\rho) = \ln Z_{MF} - \ln Z + \beta \langle H_{MF} - H \rangle_{H} \; .
\end{equation}
This is a very convenient expression since it reduces the total amount of entanglement
in the state $\rho$ to calculations of the energy and free energy of the state and its
mean field approximation. Moreover, by differentiating twice with respect to $B$ we
can conclude that
\begin{equation}
\xi_{Sep} - \xi = \frac{\partial^2 E(\rho)}{\partial B^2} + \beta \frac{\partial^2
\langle H_{MF} - H \rangle_{H}}{\partial B^2} \; .
\end{equation}
Therefore the difference in the susceptibilities between our approximation and the
true value is directly proportional to (the second derivative of) the amount of
entanglement in the state $\rho$. It is thus clear that in general this entanglement
will have macroscopic effect on quantities such as the susceptibility (as well as
other quantities). This offers a more general and formal theoretical justification for
the observations in \cite{Nature}.

Although this equality is very suggestive from the conceptual perspective, it is most
likely useless from the practical point of view of computing the amount of
entanglement. This is because it is very difficult to compute the form of the closest
separable state to a given entangled state. What is possible, thought, and can be very
useful as well, is to calculate the mean field approximation.

So in general our approximation to $\rho$ is not likely to be the closest separable
state, but some completely uncorrelated state instead. Then we have the following
inequality:
\begin{equation}
E(\rho) \le \ln Z_{MF} - \ln Z + \beta \langle H_{MF} - H \rangle_{H}\; .
\label{upperbound}
\end{equation}
This is an upper bound on the amount of entanglement whatever $\rho_{MF}$ we choose.
In fact, the right hand side will more appropriately be describing the total
multi-partite mutual information. We would like to illustrate the usefulness of this
formalism using the one dimensional Ising Model in a Transverse Field. Before that, we
analyze a much simpler model of two qubits interacting through a Heisenberg coupling.

\section{Simple example: Two qubit Antiferromagnetic Heisenberg Model}

Thermal entanglement properties of two qubits coupled through a Heisenberg interaction
are very easily understood. The Hamiltonian for the 1D Heisenberg chain in a constant
external magnetic field $B$, is given by
\begin{equation}
H=\sum_{i=1}^2 (B \sigma_i^z + \frac{J}{2}
\vec{\sigma}_i.\vec{\sigma}_{i+1})
\end{equation}
where $\vec{\sigma}_i=(\sigma_i^x, \sigma_i^y, \sigma_i^z)$ in
which $\sigma^{x/y/z}_i$ are the Pauli matrices for the $i$th
spin. The regimes $J>0$ and $J<0$ correspond to the
antiferromagnetic and the ferromagnetic cases respectively. The
state of the above system at thermal equilibrium (temperature $T$)
is $\rho=e^{-H/kT}/Z$ where, as before, $Z$ is the partition
function and $k$ is Boltzmann's constant. The factor half in $J/2$
is there for consistency reasons with the later notation.

We first examine the $2$ qubit antiferromagnetic chain. Since this is only a two qubit
mixed state we can compute entanglement of formation exactly. We can also compute the
mutual information exactly and so, there is no need to use any approximate methods.
However, we will use the bound previously derived just to illustrate the method and
compare it to the exact values. Before that, we review the exact results following the
paper by Arnesen et al \cite{ved01}.

For $B=0$, the singlet is the ground state and the triplets are the degenerate excited
states. In this case, the maximum entanglement is at $T=0$ and it decreases with $T$
due to mixing of the triplets with the singlet. For a higher value of $B$, however,
the triplet states split, and $|00\rangle$ becomes the ground state. In that case
there is no entanglement at $T=0$, but increasing $T$ increases entanglement by
bringing in some singlet component into the mixture. On the other hand, as $B$ is
increased at $T=0$, the entanglement vanishes suddenly as $B$ crosses a critical value
of $B_c=2J$ when $|00\rangle$ becomes the ground state. This special point $T=0,
B=B_c$, at which entanglement undergoes a sudden change with variation of $B$, is the
point of a {\em quantum phase transition} (phase transitions taking place at zero
temperature due to variation of interaction terms in the Hamiltonian of a system).
Note that this is, strictly speaking, not a real phase transition which can only
appear in the infinite $N$ limit. This is because for any finite $N$ entanglement
function (as well as all the other thermodynamic functions) are analytic. At any
finite $T$, however, entanglement decays off analytically after $B$ crosses $B_c$. In
the ferromagnetic case, the state of the system at $B=0$ and $T=0$ is an equal mixture
of the three triplet states. This state is disentangled. Increasing $B$ increases the
proportion of $|00\rangle$ in the state which cannot make it entangled. Increasing $T$
increases the proportion of singlet in the state which can only decrease entanglement
by mixing with the triplet. Thus we never find any entanglement in the $2-$qubit
ferromagnet. These features of the $2-$qubit Heisenberg model are also present in the
$N$ qubit model \cite{ved01}.

Let us now apply the bound previously derive and see how well it
agrees with the exact results. This is, of course, just an
exercise in computing and testing the bound rather than having any
practical purpose. It is convenient to do it simply because we
have all the results we need and the model is really very easy to
understand. On the other hand, it captures some important features
of more complicated models and, therefore, it will prepare us for
the next section. For us of great importance is that the
eigenvalues of the thermal state are easily found to be $2B+J/2,
J/2, -2B+J/2, -3J/2$, the first three corresponding to the triplet
states and the last to the singlet state. The partition function
is then computed to be:
\begin{equation}
Z = 2(e^{-K} \cosh 2C + e^{K}\cosh 2K)\; .
\end{equation}
In order to compute the bound, the second step is to establish the mean field
Hamiltonian. One possibility is the Hamiltonian of the form
\begin{equation}
H_{MF} = (B + \frac{J}{2}s_z) \sigma_z + s_x \frac{J}{2} \sigma_x
+ s_y \frac{J}{2} \sigma_y
\end{equation}
for each of the two spins. The total Hamiltonian is just the sum of the above for each
of the spins, and the values $s_{x,y,z}$ are the mean field average values of the spin
in the $x,y$ and $z$ direction. The eigenvalues of the mean field Hamiltonian are $\pm
\sqrt{(B+s_zJ)^2 + (s^2_x + s^2_y)J^2}$. The corresponding partition function is then
also easily computed to be
\begin{equation}
Z_{MF} = (2\cosh \sqrt{(C+s_zK)^2 + (s^2_x + s^2_y)K^2})^2 \; .
\end{equation}
The ultimate square in the expression comes from the fact that
there are two spins and each has the same identical individual
partition function. The final quantities we have to calculate are
the averages of the two Hamiltonians. We obtain for the average of
$H$,
\begin{equation}
\beta \langle H\rangle_H = \frac{1}{Z} \{(2C+K)e^{-(2C+K)} + Ke^{-K} +
(-2C+K)e^{-(-2C+K)}+ (-3K)e^{3K}\}\; .
\end{equation}
On the other hand, we can also calculate the average mean-filed energy to be
\begin{equation}
\beta \langle H_{MF}\rangle_H = \frac{-1}{Z} \{2(C+Ks_z)e^{-K} \sinh (2C) \} \; .
\end{equation}
We now need an estimate for the average spin values $s_{x,y,z}$. This is done by
computing the averages of the corresponding Pauli operators for the mean-filed state.
Strictly speaking, the averages should be computed with respect to the original state
$\rho$, but this is in practice difficult to handle and it is the reason for using the
mean field approximation in the first place. Thus we have to settle for computing the
following averages $\langle \sigma_x\rangle_{MF} = \langle \sigma_y\rangle_{MF} = 0$
and
\begin{equation}
s_z = \langle \sigma_z \rangle_{MF} = - \frac{e^{-K} \sinh (2C)}{2(e^{-K} \cosh 2C +
e^{K}\cosh 2K)} \; .
\end{equation}
Therefore only the average in the $z$ direction is non-zero. Here we are somewhat
unusually lucky and there is no $s_z$ variable on the right hand side. Therefore,
given $K$ and $C$, the corresponding $s_z$ is very easy to calculate. This is in
general not the case as we have seen in eq. (\ref{mean field}). We have plotted the
dependence of $s$ on $k$ and $C$ in Fig. 1. As we can see that value is always less
than zero and never less than $-1$ which is a reasonable behaviour.

\begin{figure}[ht]
\begin{center}
\hspace{0mm} \epsfxsize=7.0cm
\epsfbox{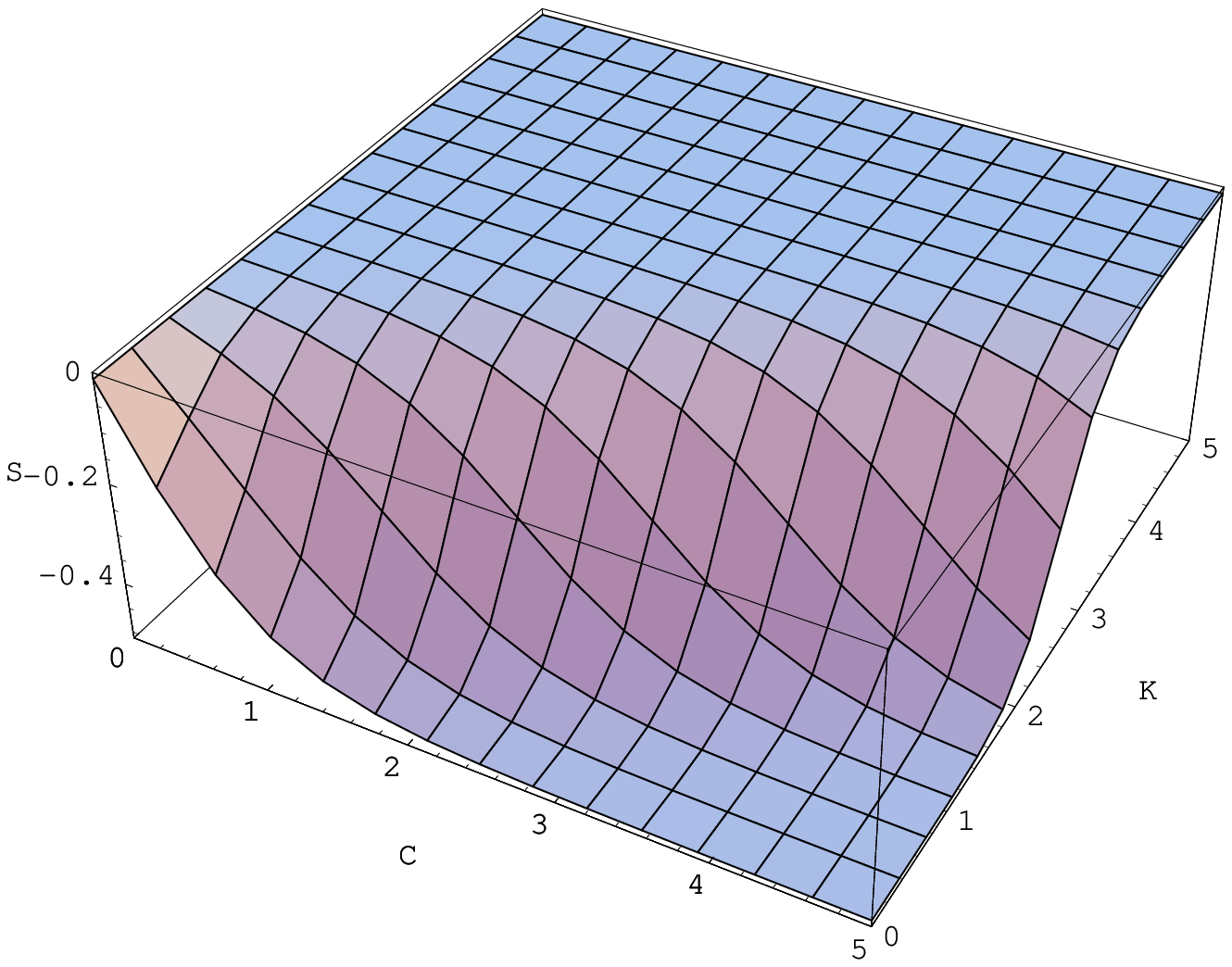}\\[0.2cm]
\begin{caption}
{
This figure shows the dependence of the mean field value of $s$ on $K$ and $C$. Note
that there is a transition beyond the line $2K=C$, for which the value of
magnetisation becomes zero. This will be reflected in the behaviour of the bound on
mutual information as will be seen in the next figure.}
\end{caption}
\end{center}
\end{figure}

Putting all these results together, the upper bound on correlations (entanglement) in
eq. (\ref{upperbound}) is now given by
\begin{eqnarray}
E(\rho) & \le & \ln (2\cosh (C  - \frac{e^{-K} \sinh (2C)}{2(e^{-K} \cosh 2C +
e^{-K}\cosh 2K)} K))^2 - \ln (2(e^{-K} \cosh 2C +
e^{-K}\cosh 2K)) \nonumber \\
& + & \frac{1}{Z} \{(2C+K)e^{-(2C+K)} + Ke^{-K} +
(-2C+K)e^{-(-2C+K)}+ (-3K)e^{3K}\} \nonumber \\
& - & \frac{1}{Z} \{2(C  - \frac{e^{-K} \sinh (2C)}{2(e^{-K} \cosh 2C + e^{-K}\cosh
2K)}K)e^{-K} \sinh (2C) \} \; .
\end{eqnarray}
We have plotted this bound in Fig. 2. We can see that it has reasonable behaviour, in
the sense that, for example, it increases with $K$ as we expect correlations to
increase with increasing interaction. Note that there is an increase in the value of
the bound when $2K>C$. This is interesting when compared with the behaviour of
entanglement. Like we said before, entanglement also undergoes a transformation from a
non-zero value to a zero value at this point as argued before \cite{ved01}. We can
also see some unreasonable behaviour from our approximation: the bound can be larger
than $2$, but the mutual information between two qubits can never exceed this amount.
For some values of $K$ and $C$, therefore, the bound becomes trivial (it is not wrong
strictly speaking, but it just becomes useless). Note that for $K=0$ the bound also
becomes zero, and hence there are no correlations between the spins. This is to be
expected as $K$ signifies the strength of the interaction and the absence of
interaction means that the spins cannot become correlated in any way. For this case,
our mean field approximation becomes exact.

\begin{figure}[ht]
\begin{center}
\hspace{0mm} \epsfxsize=7.0cm
\epsfbox{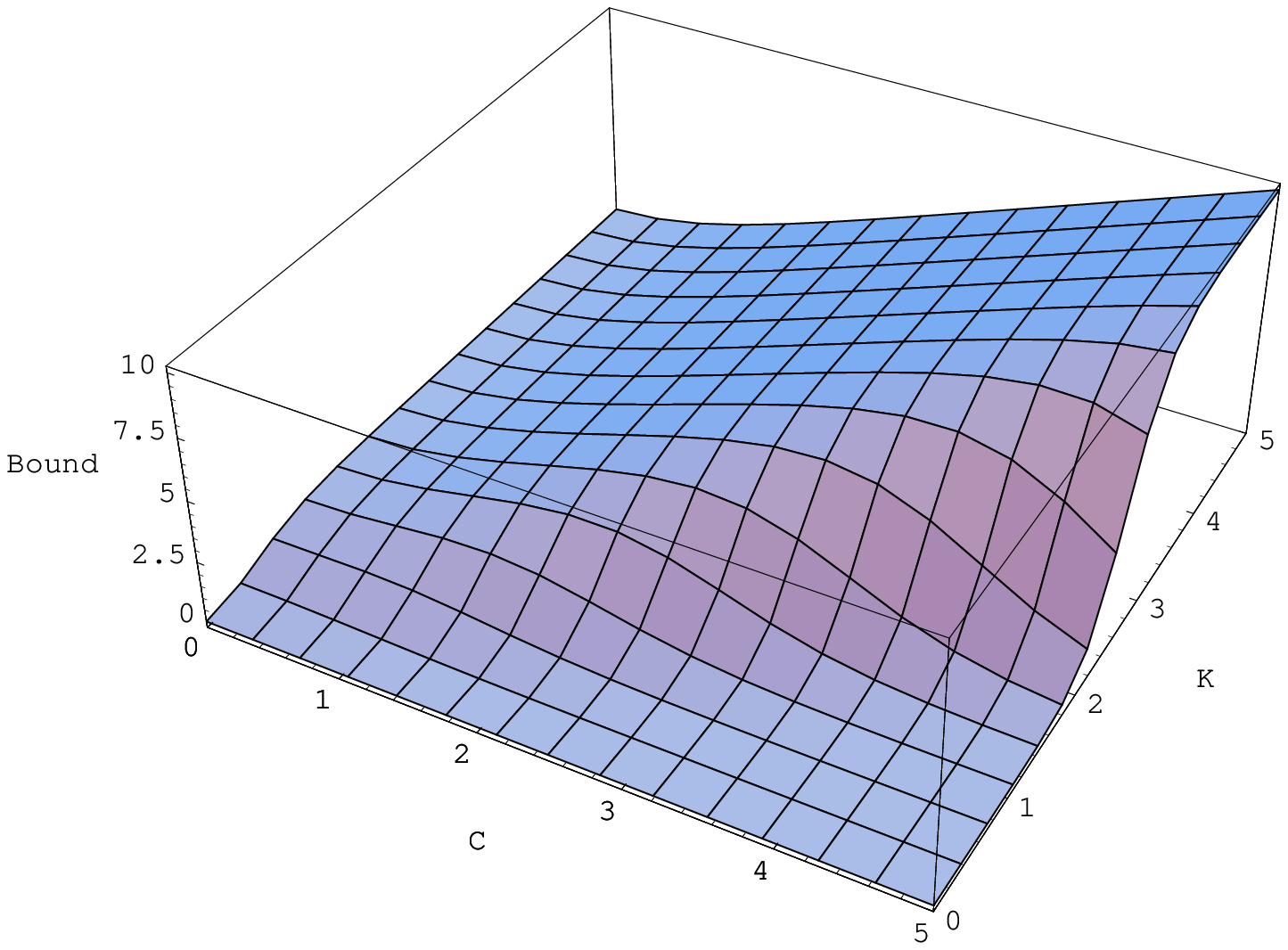}\\[0.2cm]
\begin{caption}
{
This figure shows the dependence of our bound to the amount of correlations for two
Heisenber spins as a function of $K$ and $C$. Note the transition at $2K=C$ and the
disappearance of correlations for $K=0$.}
\end{caption}
\end{center}
\end{figure}

We can compare the bound to the true value of mutual information (which should be
lower unless the mean-filed approximation leads exactly to the closest product
state!). This is just the difference of the sum of individual qubit entropies (which
are equal to each other in this case) and the total two-qubit entropy, $I = 2S(\rho_1)
- S(\rho)$. It is given by
\begin{eqnarray}
I & = &  \frac{e^{-(2C+K)}}{Z} \ln \frac{e^{-(2C+K)}}{Z} + \frac{e^{-
K}}{Z} \ln \frac{e^{-K}}{Z}  \nonumber \\
& + & \frac{e^{-(-2C+K)}}{Z} \ln \frac{e^{-(-2C+K)}}{Z} + \frac{e^{3
K}}{Z} \ln \frac{e^{3 K}}{Z} \nonumber \\
& - & 2 \{\frac{e^{-K}/2 + e^{3K}/2 + e^{-(2C+ K)}}{Z} \ln \{
\frac{e^{-K}/2 + e^{3K}/2 + e^{-(2C+ K)}}{Z} \} \nonumber \\
& + & \frac{e^{-K}/2 + e^{3K}/2 + e^{-(-2C+ K)}}{Z} \ln \{ \frac{e^{-K}/2 + e^{3K}/2 +
e^{-(-2C+ K)}}{Z} \} \} \; .
\end{eqnarray}
This function is plotted in Fig 3. Although the behaviour is different, the same
transition occurs at $2K = C$.

\begin{figure}[ht]
\begin{center}
\hspace{0mm} \epsfxsize=7.0cm
\epsfbox{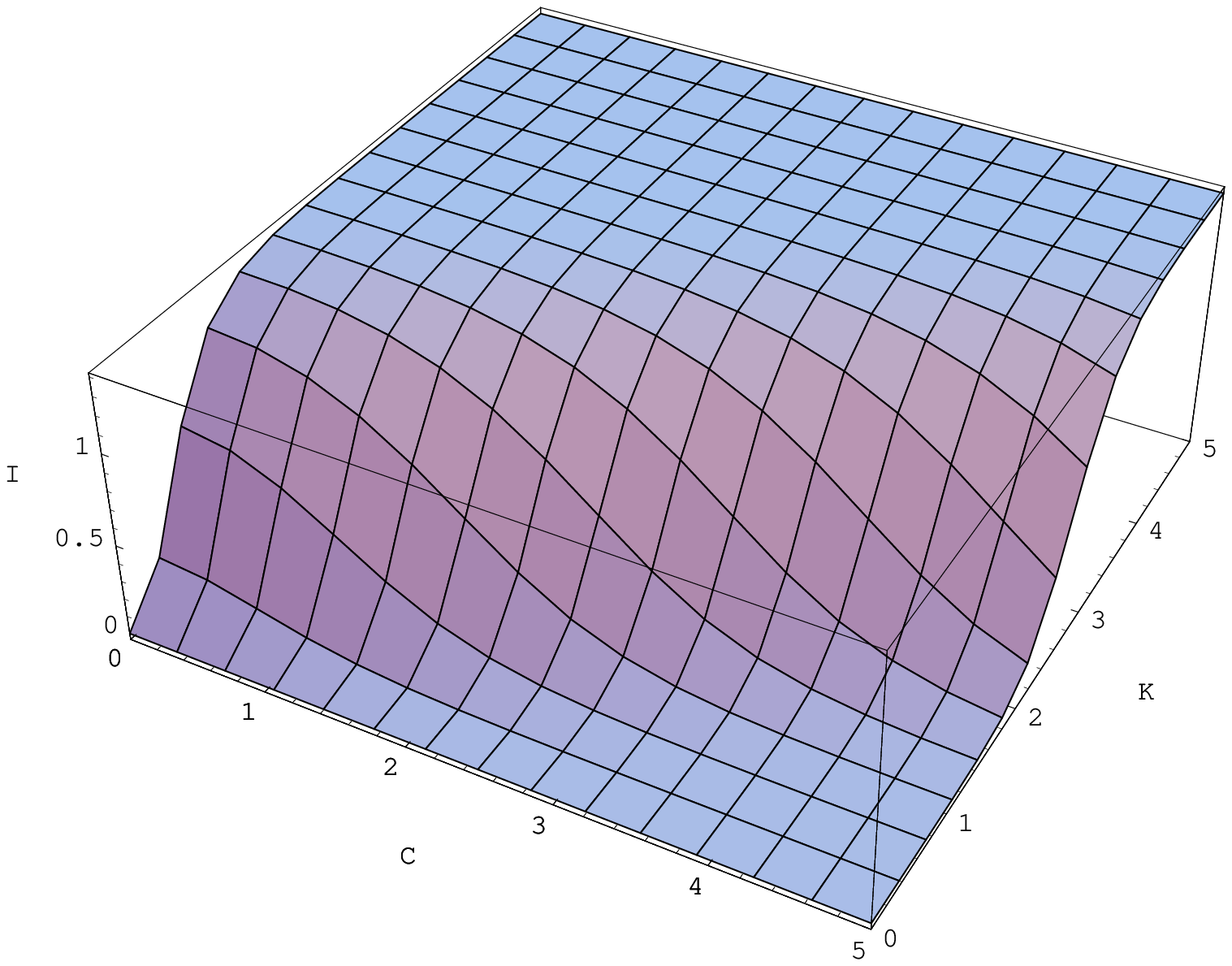}\\[0.2cm]
\begin{caption}
{
This figure shows the exact dependence of mutual information for the two spin
Heisenberg model on the values of $K$ and $C$. As noted in the text, there are no
correlations in the absence of the interaction $K$.}
\end{caption}
\end{center}
\end{figure}

We can see that our bound is always greater than this value (as it should be!). Fig 4
shows the difference between the bound on correlations and the actual mutual
information. Note that for $K=0$ and in the vicinity of this region the bound still
gives exact results.

\begin{figure}[ht]
\begin{center}
\hspace{0mm} \epsfxsize=7.0cm
\epsfbox{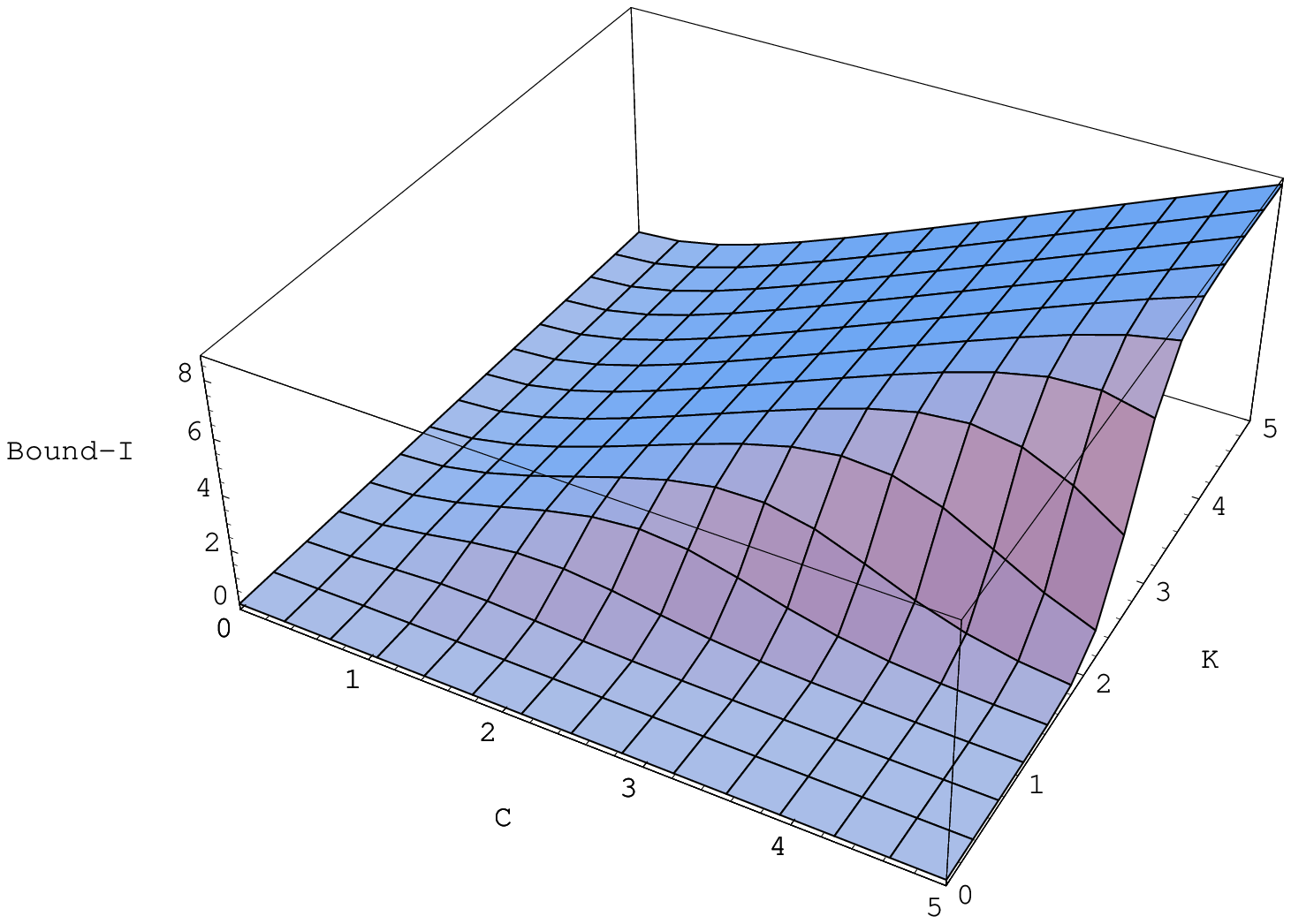}\\[0.2cm]
\begin{caption}
{
This figure shows the difference between the bound and the exact value of $I$ as a
function of $K$ and $C$.}
\end{caption}
\end{center}
\end{figure}

We are now in a position to apply our ideas to a much more complex model, whose
features are still being investigated \cite{sach99}.

\section{One Dimensional Ising Model in a Transverse
Field}

This model has been extensively used in many situations and has been analyzed in
various regimes \cite{sach99}. The interesting feature of the Hamiltonian which makes
it different from the ordinary classical Hamiltonian is that the interaction part of
the Hailtonian and the external field point in different mutually orthogonal
directions. This makes the model more complex than if the two parts of the Hamltonian
were in the same direction (as in the original Ising model). One of the main
consequences is that while there is never any entanglement present in the original
Hamiltonian model, the one in the transverse field can exhibit entanglement
\cite{ved02}.

The one dimensional Ising Model in a Transverse Field is completely specified by the
following Hamiltonian:
\begin{equation}
H = -\frac{J}{2} \sum_{i=1}^{N} \sigma_i^x\otimes \sigma_{i+1}^x - B \sum_{i=1}^{N}
\sigma_i^z \; .
\end{equation}
We first briefly summarize how to diagonalize this Hamiltonian \cite{Kat}. The main
tools are the Jordan-Wigner transformation, followed by a Fourier transform and
finally followed by the Bogoliubov transform. What this achieves is the following: the
initial spins, which are localized fermions, are subsequently converted into highly
non-local fermions and then the delocalized fermionic Hamiltonian is easily
diagonalised to find its spectrum. To simplify mathematics even further at this point
we will divide the Hamiltonian through by $J/2$ and use as the only variable $\lambda
= 2B/J$, so that
\begin{equation}
H = -\sum_{i=1}^{N} \sigma_i^x\otimes \sigma_{i+1}^x - \lambda \sum_{i=1}^{N}
\sigma_i^z \; .
\end{equation}
We will go back to the original Hamiltonian, recovering $B$ and
$J$ at the end when we present results of our calculations. The
spin operators in this expression satisfy anticommutation rules at
any given site (just like fermions) but follow commutation rules
at separate sites (again just like fermions). The non-local
Jordan-Wigner transformation maps these operators into fully
anticommuting spinless fermions defined by
\begin{equation}
a_l =  \prod_{m<l}\sigma_m^z \frac{\sigma_l^x +
 i\sigma_l^y}{2}
\end{equation}
such that
\begin{equation}
\{a_l^{\dagger}, a_m\}=\delta_{lm} , ~~~\{a_l, a_m\} =0 \; .
\end{equation}
In terms of operators $a$ the above Hamiltonian becomes
\begin{equation}
H = \frac{1}{2} \sum_{l=-\frac{N}{2}+1}^{\frac{N}{2}-1}
 \{ \left(a_{l+1}^\dagger a_{l} + a_l^\dagger a_{l+1}\right)
 + \left(a_l^\dagger a_{l+1}^\dagger + a_{l+1} a_l
 \right)\}
 - \lambda\!\!\! \sum_{l=-\frac{N}{2}+1}^{\frac{N}{2}}
 \!\! a_l^{\dagger} a_l \; .
\end{equation}

We now introduce the Fourier transformed (fermionic) operators \be d_k
=\frac{1}{\sqrt{N}} \sum_{l=-\frac{N}{2}+1}^{\frac{N}{2}-1} a_l
e^{-i\frac{2\pi}{N}kl}, \ee $-N/2+1 \leq k \leq N/2$. Due to the fact that this
transformation is unitary, the anticommutation relations remain true for the new
$d$-operators. The Hamiltonian now takes an almost diagonal form,
\begin{equation}
H=\sum_{k=-N/2+1}^{N/2} (-\lambda+\cos \frac{2\pi k}{N}) \, d^{\dagger}_k d_k +
\frac{i}{2} \sum_{k=-N/2+1}^{N/2} \sin\frac{2\pi k}{N} \, ( d_k d_{-k} + d^{\dagger}_k
d^{\dagger}_{-k}) \; .
\end{equation}
where an extra term, suppressed by $\frac{1}{N}$, should be
present. This terms clearly tends to zero as $N\to \infty$ and we
can therefore safely omit it as all our results will refer to the
thermodynamic limit.

A final unitary transformation is now necessary to cast the
Hamiltonian into a manifestly diagonal form. This so-called
Bogoliubov transformation can be expressed as
\begin{equation}
b^{\dagger}_k =u_k \, d^{\dagger}_k + i v_k \, d_{-k} b_k = u_k \, d_k - i v_k \,
d^{\dagger}_{-k}
\end{equation}
where $u_k = \cos{\theta_k/2},~ v_k = \sin{\theta_k/2}$ for
$\cos{\theta_k} = \frac{-\lambda+\cos{\frac{2\pi}{N}k}}
{\sqrt{(\lambda-\cos{\frac{2\pi k}{N}})^2+\sin^2{\frac{2\pi
k}{N}}}}$.  Again, due to unitarity of the Bogoliubov
transformation the $b$-operators follow the usual anticommutation
relations. Finally, the Hamiltonian takes the following diagonal
form
\begin{equation}
H=\sum_{k=-N/2+1}^{N/2} \tilde{\Lambda}_k
 \, b^{\dagger}_k b_k
\end{equation}
where
\begin{equation}
\tilde{\Lambda}_k \equiv \sqrt{\left(\lambda-\cos{\frac{2\pi
k}{N}}\right)^2+\sin^2{\frac{2\pi k}{N}}} \; .
\end{equation}
The thermodynamical limit is obtained by defining $ \phi =2\pi k/N$ and taking the
$N\to \infty$ limit
\begin{equation}
H = \int^{\pi}_{-\pi}\frac{{\rm d}\phi}{2\pi}\ \Lambda_\phi b^{\dagger}_\phi b_\phi,
\end{equation}
with
\begin{equation}
\Lambda^2_\phi = (\lambda-\cos{\phi})^2 + \sin^2{\phi} = K^2 + C^2 -2KC\cos \phi \; .
\end{equation}
We have in the last step reintroduced the $K$ and the $C$
parameter in order to recover the field and interaction dependence
which will be important in our subsequent analysis. The exact
partition function can from here immediately be calculated (this
was originally done by Katsura \cite{Kat}). In the large $N$ limit
(which is what we are always interested in when it comes to
computing thermodynamical quantities) we have
\begin{equation}
\ln Z = N\ln 2 + N \frac{1}{\pi} \int_0^{\pi} \ln \{\cosh \sqrt{K^2 + C^2 - 2KC\cos
\omega}\} d\omega \; .
\end{equation}

We now need to construct a mean field Hamiltonian. Let us try to approximate this
Hamiltonian with the following one:
\begin{equation}
H_{MF} = \sum_l (-\frac{1}{2} sJ \sigma_l^x - B \sigma_l^z) \; .
\end{equation}
This Hamiltonian is completely local and its eigenstates cannot be entangled (not even
classically correlated). The parameter $s$ is meant to approximate the effect of all
other spins on the spin $l$. We will at the moment leave the value of $s$
undetermined. The eigenvalues of $-\frac{1}{2} sJ \sigma_l^x - mB \sigma_l^z$ are
easily computed to be:
\begin{equation}
\lambda_{\pm} = \pm \sqrt{B^2 + \frac{1}{4}s^2J^2}\; .
\end{equation}
The mean field partition function is, therefore, given by
\begin{equation}
Z_{MF} = 2^N (\cosh \sqrt{C^2 + s^2K^2})^N
\end{equation}
where, as before, $C= B/kT$ and $K = J/2kT$. The $N$-th power again reflects the fact
that there are $N$ spins each of which has the same identical partition function. The
free energy is in the mean field theory now proportional to
\begin{equation}
\ln Z_{MF} = N \ln 2 + N \ln \{\cosh \sqrt{C^2 + s^2K^2}\}
\end{equation}
In order to complete the upper bound on thermal entanglement  for the whole chain we
need to estimate the average of the difference between the mean filed Hamiltonian and
the true Ising Hamiltonian. The average of $H$ is known from the partition function
and it is \cite{Kat}
\begin{equation}
\langle H\rangle_{H} = -NkT \frac{1}{\pi} \int_0^{\pi} \sqrt{K^2 + C^2 - 2KC\cos
\omega} \tanh \sqrt{K^2 + C^2 - 2KC\cos \omega} d\omega \; .
\end{equation}
The final quantity left to calculate is the average of the mean field Hamiltonian
which requires us to compute the averages of $\sigma_z$ and $\sigma_x$. The average of
the former is known and it is
\begin{equation}
\langle \sigma_z\rangle_H = Nm \frac{1}{\pi} \int_0^{\pi} \frac{(C-K\cos\omega)\tanh
\sqrt{K^2 + C^2 - 2KC\cos \omega}}{\sqrt{K^2 + C^2 - 2KC\cos \omega}}d\omega \; .
\end{equation}
The biggest problem presents now the calculation of the average of
the $\sigma_x$ operator. This value is not known analytically
unless $\beta \rightarrow \infty$. So, instead of calculating this
value exactly, we will assume that $\langle \sigma_x \rangle = s$.

Putting all the results so far together we obtain the following
bound on the amount on entanglement
\begin{eqnarray}
E(\rho) & \le N & \{ \ln \{\cosh \sqrt{C^2 + s^2K^2} -
\frac{1}{\pi} \int_0^{\pi} \ln \{\cosh \sqrt{K^2 + C^2 - 2KC\cos
\omega}\}
d\omega \nonumber \\
& +& \frac{1}{\pi} \int_0^{\pi} \sqrt{K^2 + C^2 - 2KC\cos \omega}
\tanh \sqrt{K^2 + C^2 - 2KC\cos \omega} d\omega - Ks^2 \nonumber\\
& - & C \frac{1}{\pi} \int_0^{\pi} \frac{(C-K\cos\omega)\tanh \sqrt{K^2 + C^2 -
2KC\cos \omega}}{\sqrt{K^2 + C^2 - 2KC\cos \omega}}d\omega\} \; .
\end{eqnarray}

We now turn our attention to calculating the value of $s$ which is necessary in order
to be able to used the above bound. Again, we cannot calculate the real value of $s$
in the state $\rho$, but need to simplify and assume that $s = \langle
\sigma_x\rangle_{H_{MF}}$. This means that we only need to calculate the average of
$\sigma_x$ in the mean field and this is a simple task. For this we need the
eigenvalues and eigenvectors of the mean field Hamiltonian. The former have already
been used, and the latter are (in general) given by
\begin{eqnarray}
|\psi_{+}\rangle & =  & a |0\rangle + b|1\rangle \\
|\psi_{-}\rangle & =  & b |0\rangle - a|1\rangle
\end{eqnarray}
($a$ and $b$ are real), where it can be calculated that
\begin{equation}
ab = - \frac{K(C+\sqrt{C^2 + s^2K^2})}{K^2 + (C+\sqrt{C^2 +
s^2K^2})^2}
\end{equation}
(it turns out that we will only ever need the value of the product
of $a$ and $b$ and not their individual values). The average value
of the Pauli $x$ operator is now:
\begin{equation}
s = \langle \sigma_x \rangle_{H_{MF}} = \frac{2K(C+\sqrt{C^2 +
s^2K^2})}{K^2 + (C+\sqrt{C^2 + s^2K^2})^2} \tanh \sqrt{C^2 +
s^2K^2}
\end{equation}
and this therefore presents the equation from which we should
determine the value of $s$. The algorithm for computing the upper
bound on the total amount of entanglement is now the following:
\begin{itemize}
\item Given the values of $K$ and $C$, first compute the
corresponding value of $s  = s(K,C)$; \item Then compute the value
of the upper bound for the same three values of $K,C$ and $s$.
\end{itemize}

We can, of course, always handle this problem numerically and some numerical results
will indeed be presented. We will first discuss estimating the critical temperature
beyond which we do not have the solution for the average magnetisation in the mean
field approximation. Even thought this is impossible to do analytically (we have a
transcendental equation), we can always look at some special cases. For example, when
the external field vanishes, $C=0$, we have that
\begin{equation}
\tanh Ks = \frac{1+s^2}{2}
\end{equation}
The solutions of this equation exist only if $K> K_c$ which implies that $T<J/2kK_c$,
$k$ being the Boltzmann constant. In this case, we can numerically estimate that $K_c
\approx 1.37$. Note that if, on the other hand, $K=0$, then it follows immediately
that $s=0$. It is also not so difficult to obtain other critical regions. For example,
for $C=1$, values of $K\ge 8$ all yield the value of $s$ outside the acceptable region
(which is between $-1$ and $1$). In general, therefore, it is first important to
investigate the domain of validity of the approximation before we can use the bound on
correlations.

Before we analyse the general values of the interaction, external
field and temperature, let us first investigate a few special
cases which can be addressed immediately and analytically. First
of all, as $K$ tends to zero (while $C$ is kept finite), the value
of the upper bound tends to zero. Therefore, there is no
entanglement present here. This is expected as the small $K$ limit
implies that the external field dominates the interaction part and
then the eigenstates become product states of all spins pointing
in the $z$ direction which is a completely disentangled state
(this is why the average magnetization in the $x$ direction, $s$,
tends to zero as it is equally likely to point in the positive and
negative directions).

The opposite limit is when $K$ becomes large (while $C$ is still
kept finite). Then, as the interaction dominates the external
field, the state of the system approaches the product of spins all
aligned and pointing in the $x$ direction (the ground state is
when all the spins assume the same eigenvalue of $x$, the first
excited states are the ones where one spin has the opposite $x$
value to the rest, and so on). Therefore, we should expect no
entanglement in this case, although it may seem surprising that a
strong interaction ultimately does not produce entangled states.
From the bound we obtain
\begin{equation}
\frac{E(\rho)}{N}\le K(s-s^2)
\end{equation}
The mean field theory also implies that $s\rightarrow 1$, and so entanglement indeed
disappears as the bound tends to zero.

The third special case we analyze is when $C=K \rightarrow \infty$. There are two
different physical ways we can reach this limit. One is the keep the temperature
constant and increase the external field and the interaction strength at the same
rate. The other is to keep $B$ and $J$ fixed and decrease the temperature to zero.
From the mean field approximation we can obtain the value of $s\approx 0.75$. This
then gives us the bound on entanglement of
\begin{equation}
\frac{E}{N} \le \{\frac{11}{16} - \frac{1}{\pi} \int_0^{\pi} \sqrt{1-\cos \omega} d
\omega \} \approx 0.2 K
\end{equation}
The regime $T=0$ has been extensively studied \cite{sach99} and, as we discussed
before, there is a quantum critical point of $B$ at which some thermodynamic
potentials (such as the free energy) become non-analytic. At $T=0$, the critical
region is when $B < J/2$ \cite{sach99} (in other words this is the $C<K$ region, but
both in the large limit. This behaviour is confirmed by our calculation as will be
seen when we present plots of the bound. Note, however, that the upper bound on
information per spin grows linearly with $K$. Since the mutual information per spin
cannot be larger than the entropy of the spin, which is maximally equal to $\ln 2$,
this means that the bound becomes useless once $K$ is greater than $5$. This number
has only approximately been calculated and in the limit of large $K$, nevertheless,
there will still be a cut-off of this kind beyond which our approximation is no longer
useful (although it is still, of course, valid, but only trivially).

In the case of the Heisenberg model with two qubits we compared
the bound to the exact value of the mutual information. Can we do
the same here? The answer is that in general this is not possible.
Some special cases can be addressed (such as the zero temperature,
ground state entanglement and entropies as in \cite{vid02}), but
for any generic values of $C,K,T$ it seems to be very difficult to
obtain the single spin density matrix. In particular, while we are
able to compute the average value of $\sigma_z$, there is no way
of computing the $x$ and the $y$ components necessary to fully
infer the state. This therefore prevents us from calculating the
total mutual information which would be equal to $I = N S(\rho_1)
- S(\rho)$. This is why the method presented here may be useful in
estimating correlations. Note that the total entropy $S(\rho)$, on
the other hand, can be computed, as we know the spectrum exactly.

Now we will plot the value of the upper bound for the Ising Model in a Transverse
Field for a range of values of $K$ and $C$. This is exactly what we did in the case of
two Heisenberg interacting spins. We will first reproduce the plot of the mean field
value of $s$ (which will be a transcendental equation in this case, and hence not
analytically solvable) and then use this value to plot the upper bound on
correlations. The numerical results here will also shows us in which region the
approximation can be trusted and where it definitely fails us.

The figure 5 shows the dependence of the mean field magnetization on the external
field and the coupling. One region of failure of the mean field approximation for this
model can immediately be seen from the plot. Note that for $C=1$ we have that there
are some values of $s$ which are higher than one (in general, we have that $|s|\le
1$). This is the sufficient condition for failure, but not necessary. There are other
regions where the approximation fails - it is only an approximation after all, but
they will not concern us at present.

\begin{figure}[ht]
\begin{center}
\hspace{0mm} \epsfxsize=7.0cm
\epsfbox{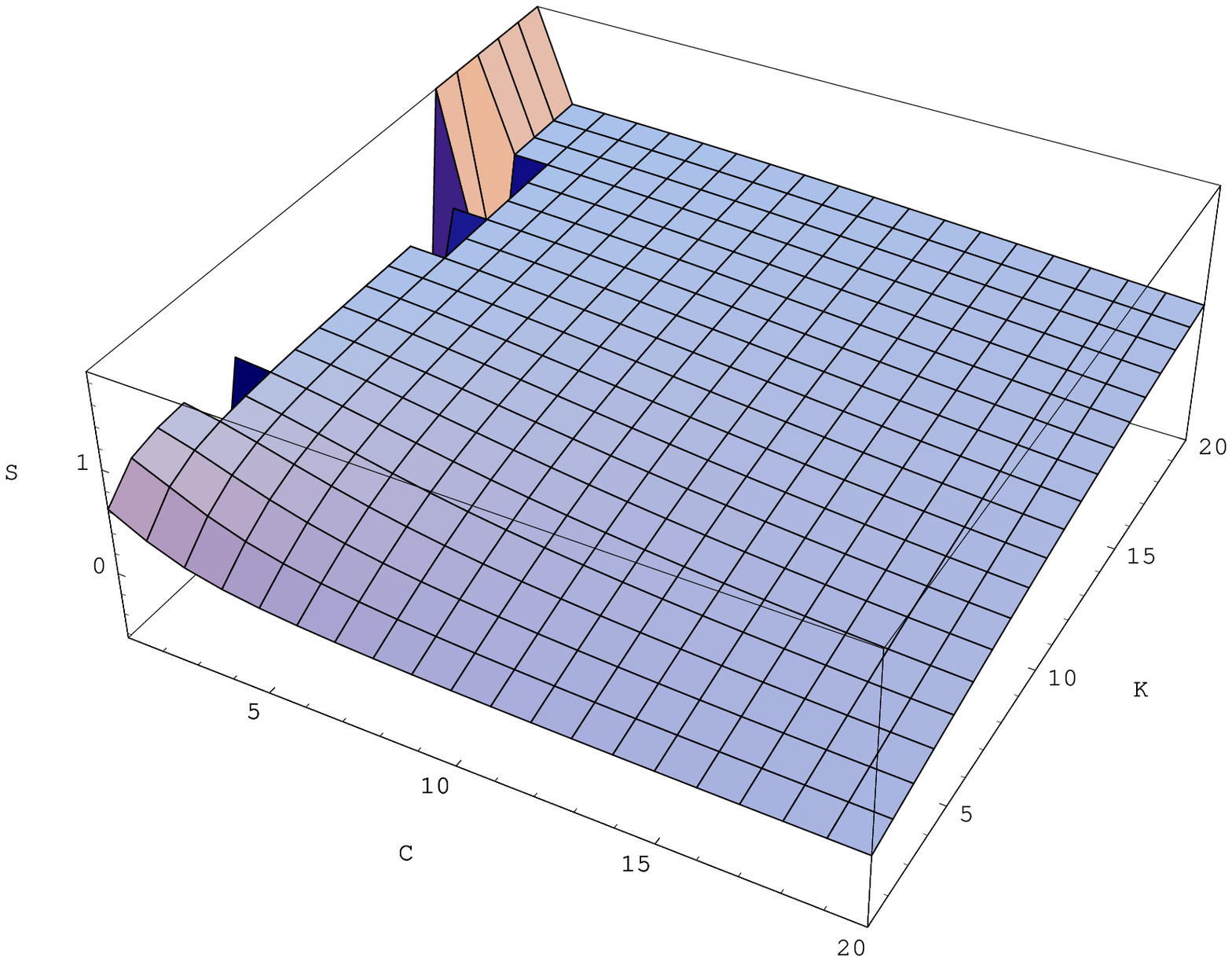}\\[0.2cm]
\begin{caption}
{
This figure shows the dependence of our mean field magnetization $s$ on the parameters
$K$ and $C$.}
\end{caption}
\end{center}
\end{figure}

With this in mind, we can now go ahead and plot the upper bound using the results for
the magnetization from figure 5. The figure 6 contains the plot of the upper bound.
The first thing that strikes us is the existence of two different regions, separated
by the boundary where $K=C$. This, very interestingly, coincides with the existence of
the critical region, $K>C$, known as the quantum criticality as introduced by Sachdev
\cite{sach99}. We have already remarked that the quantum criticality is only
manifested at zero temperature, and by varying the strength of the external field. In
our case, this coincides with the domain when $K,C \rightarrow \infty$. In our bound,
the separation between regions exists, on the other hand, for any values of $K$ and
$C$. There is no discontinuity here between the finite temperature region and the
$T=0$ region. It is not clear to the author if this finding has any more general
significance for the behaviour of the solid. Note also that there is a non-smooth
behavior for the small values of $C$, just where the value of $s$ from the mean field
theory also fails to be in the region between $\{-1,1\}$. Intuitively we expect the
approximation to fail here as we are in the region where the external field is
dominated by the interaction, and the mean-filed approximation neglects the
interaction completely.

\begin{figure}[ht]
\begin{center}
\hspace{0mm} \epsfxsize=7.0cm
\epsfbox{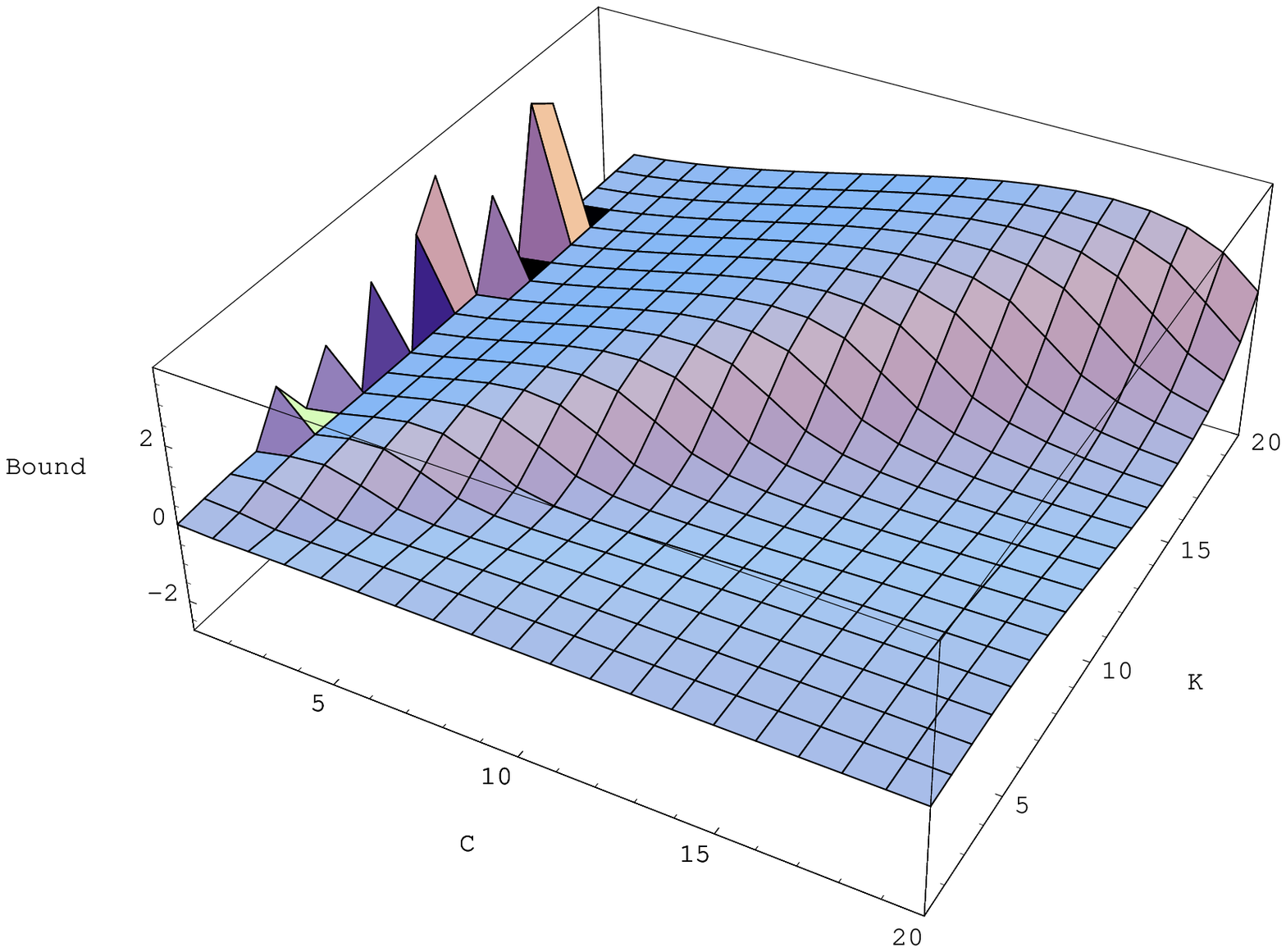}\\[0.2cm]
\begin{caption}
{
This figure shows the dependence of our correlation bound on the parameters $K$ and
$C$. Note that the region for which $K>C$ shows an increase in the bound. This
coincides with the notion of the quantum phase transition.}
\end{caption}
\end{center}
\end{figure}

Having obtained the above results, how much should we trust our method? In other
words, what is the domain of validity of the mean field approximation? First of all,
we can see when the method obviously fails. This is, as we mentioned above a number of
times, when the value of the magnitude (absolute value) of $s$ becomes greater than
$1$. Also, we have failure when the bound on correlations per spin exceeds the maximum
value of $1$. We have already commented on when this happens in our model. There is,
however, a more general way of assessing the validity of the approximation. We will
only present a sketch of this argument here, and a much more general analysis can be
found in any more advanced statistical mechanics book (e.g. \cite{Kubo}).

The usual way of deriving the limitations of the mean field approximation is to look
at the regime when fluctuations due to interactions become large. This is when we
expect the mean field to fail since the mean field neglect those interactions. How can
this regime be determined more precisely? We know that the susceptibility is given by
the following expressions
\begin{equation}
\chi = \frac{\partial \bar M}{\partial B} = \frac{\bar{M^2} - (\bar M)^2}{kT} =
\frac{\bar{\Delta M^2}}{kT}
\end{equation}
where $\bar M$ is the average magnetization. We can recast the second equality to
obtain:
\begin{equation}
\frac{\bar{\Delta M^2}}{(\bar M)^2} = \frac{kT\chi}{(\bar M)^2}
\end{equation}
For fluctuations to be small we need to have that
\begin{equation}
kT\chi << (\bar M)^2
\end{equation}
More detailed calculations of $(\bar M)^2$ \cite{Kubo} suggest that the dimensionality
of the system would have to be greater than $4$ in order for this to be correct. This
is intuitively pleasing, as higher dimensions imply that each spin has more neighbours
and therefore we expect to obtain a better mean field average. In other words, this
means that the overall state displays a low degree of correlations. Indeed, for the
mean field and the exact susceptibility to be approximately equal
\begin{equation}
\chi_{MF} - \chi \approx 0
\end{equation}
we have to have that
\begin{equation}
kT \frac{\partial^2 I}{\partial B^2} \approx \frac{\partial^2\langle
H-H_{MF}\rangle_H}{\partial B^2}
\end{equation}
And so we can see that the smaller the mutual information $I$, the better the mean
field approximation. The conclusion of the back of an envelope type of analysis is
that we should not expect the one dimensional mean field approximation to be very
accurate here. Nevertheless, the results presented in this section indicate to us that
there is a general agreement with other results obtained for this model.
Generalizations of our method to higher dimensions would certainly be most welcome.

We would now like to discuss another question of very general importance. So far, we
have been careful to point out that our bound has been limiting the amount of total
correlations - and thus entanglement, but we have not tried to discriminate between
the quantum and the classical contribution to correlations. Recently, however, we have
seen claims that the effects of entanglement on the macroscopic properties and
genuinely different from the effects of classical correlations \cite{Nature}. But, how
can we be sure that we have entanglement in a solid and that its effects cannot be
reproduced by the ordinary, classical correlations?

\section{Classical Versus Quantum Correlations}

We have so far been discussing the effect of correlations as quantified by the mutual
information on the macroscopic quantities for some simple (one dimensional) spin
models. Of course, in a real solid, there is usually a huge number of degrees of
freedom to consider and we are here approximating them only with spin half systems.
So, how confident are we that we have really identified effects of entanglement as
opposed to some kind of interaction between these huge multitude of neglected degrees
of freedom. Can we somehow reproduce the effect of entanglement by allowing more
degrees of freedom? The answer is ``yes", and so we have to be very careful under
those circumstances if we are to identify something as a clear effect of entanglement.
We will first illustrate this with a very simple example, and will then discuss a more
general model, which is extensively used in statistical mechanics, field theory and
condensed matter physics.

Suppose that we have two correlated qubits. The maximum value of entanglement is $\ln
2$, however the mutual information can be as large as $2\ln 2$. Our understanding of
this difference is that the mutual information represents the total correlations and
is therefore the sum of quantum and classical correlations (whatever they are defined
to be - see \cite{Henderson} for one possible entropic definition). This would suggest
that in a maximally entangled state of two qubits we have $\ln 2$ of classical
correlations (for two qubits we cannot have more than this worth of classical
correlations) and $\ln 2$ worth of entanglement. So, having two qubits and knowing
that $I = 2\ln 2$ immediately implied a maximal unit of entanglement. However, suppose
that these two qubits were really two $4$-level systems. Then classical correlations
could on their own be as high as $2\ln 2$. In which case, the mutual information of
$2\ln 2$ would not necessarily imply the existence of any entanglement. Therefore, the
actual dimensionality of our constituents is very important.

It is well known that $d$ dimensional quantum statistical models
are isomorphic to $d+1$ dimensional problems in classical
statistical mechanics. Let us illustrate this with a very simple
example of a quantum mechanical two-level system - a qubit. The
question we would like to ask is: if there is entanglement in the
$d$ dimensional quantum system how is that reflected in the $d+1$
dimensional classical system?

Let us review how this correspondence works. Suppose we have a
quantum system with the Hamiltonian $H$ - this is a single system
and so the dimension $d=0$. We will now show how to correspond
this to a $d=1$ dimensional classical system. The partition
function is given by
\begin{equation}
Z = tr e^{-\beta H} = \sum_i \langle \psi_i| e^{-\beta
H}|\psi_i\rangle .
\end{equation}
where $\{|\psi_i\rangle\}_1^N$ is some orthonormal basis. Invoking
the completeness relations $\sum |\psi_i\rangle\langle \psi_i| =
1$, this can be written as
\begin{eqnarray}
Z &=& \lim_{N\rightarrow \infty} \sum_i \langle \psi_i|
(1-\frac{\beta
H}{N})^N |\psi_i\rangle \nonumber \\
&=& \lim_{N\rightarrow \infty} \sum_{i_1,i_2,...i_N}\langle
\psi_{i_1}| (1-\frac{\beta H}{N})|\psi_{i_2}\rangle \times \langle
\psi_{i_2}| (1-\frac{\beta H}{N})|\psi_{i_3}\rangle ...\times
\langle \psi_{i_N}| (1-\frac{\beta H}{N})|\psi_{i_1}\rangle
\nonumber
\end{eqnarray}
This is still a quantum mechanical expression and now we will translate it into a
``classical Hamiltonian". Suppose that we require that
\begin{equation}
\langle i| (1-\frac{\beta H}{N})|j \rangle = A e^{Bij+Ci+Dj}
\label{corresp}
\end{equation}
where $i,j$ are numbers taking values $\pm 1$ and represent the two values of the spin
(note that it is now more convenient for notational purposes to label the eigenstates
as $|\pm 1\rangle$). The values of $A,B,C$ and $D$ can be inferred from this equality.
Using the right hand side of the equality (the classical Hamiltonian) we can write the
original partition function as
\begin{equation}
Z= \lim_{N\rightarrow \infty} A^N \sum_i e^{Bu_iu_{i+1}+hu_i}
\end{equation}
where $h=C+D$, where, to make the notation closer to the classical Ising model, we
have used $u_i$ instead of $i$. Note that this is now the same as a classical one
dimensional Ising-like model with nearest neighbor interactions. In fact, this method
of equating the quantum evolution in $d$ dimension to a classical statistical problem
in $d+1$ dimension has been extensively exploited (see \cite{Feynman}). There is a
calculational advantage here since it is easier to treat exponentials of numbers (on
the right hand side) than exponentials of operators (which are on the left hand side),
but we are not interested in this motivation here.

Suppose that we have a qubit whose Hamiltonian is given by
$$
H = \left(\begin{array}{cc}
E&-\Delta\\
-\Delta&-E\end{array} \right) \; ,
$$
where $E$ and $\Delta$ are just some real numbers. The solution to
the equation (\ref{corresp}) are given by
\begin{eqnarray}
A & = & (\frac{\beta \Delta}{N})^{1/2} (1-\frac{\beta^2
E^2}{N^2})^{1/4}\\
B & = & \frac{1}{4} \ln (\frac{N^2 - \beta^2 E^2}{\beta^2
\Delta^2}) \\
h & = & \frac{1}{2} \ln (\frac{N - \beta E}{N + \beta E})
\end{eqnarray}

We are now in the position to address the question at the
beginning of this section: if $\rho = e^{-\beta H}/Z$ is an
entangled state, how is that entanglement reflected in the
classical analogue?

The above treatment is suitable for a single spin system. When we have two spins, then
we need more variables in the classical model. One way of dealing with this is the
following: take a classical chain to represent each of the qubits. To simulate the
interaction Hamiltonian between the qubits creating entanglement (correlations in
general), we need two chains to be interacting. This is a straightforward application
of the above formalism. Therefore, in order to claim that some thermodynamical
property is a consequence of two qubit entanglement, we need to make sure that we
indeed have two qubits (and not a classical chain of bits simulating that qubit). What
happens if we have $N$ qubits interacting? Then we need a two dimensional classical
square lattice of interacting spins to simulate this. A good instance of this is the
$2$ dimensional classical Ising model. This, in fact, is the reason behind the fact
that the one dimensional partition function of a transverse Ising model resembles the
partition function of Onsager for the $2$ dimensional Ising model \cite{Onsager}.
Therefore, it is not trivial to show that entanglement is responsible for some
macroscopic effects, as by enlarging our system, classical correlations can also
sometimes be used to derive the same conclusion.

At the end, we would like to point out that the above method of
making the quantum-classical correspondence is by no means the
only one. Another way of representing two interacting qubits is to
to make the coefficient $B$ spin value dependent. So we can say
that
\begin{equation}
\langle i| (1-\frac{\beta H}{N})|j \rangle = A e^{B_{ij}ij+Ci+Dj}
\end{equation}
where $B_{ij}$ are the spin-value dependent coefficients. Either way, we see that
entanglement in one dimension can always be (at least in principle) interpreted as
just classical correlations in a higher dimension. The significance that this
quantum-classical correspondence may have for our purposes, if any, is ultimately
completely unclear. What is clear, however, is that we have to be very careful to
interpret something as an effect of entanglement unless we are sure that the extra
dimensions do not contribute the extra (purely classical) correlations. One way of
being sure that entanglement is present is to perform some form of Bell inequalities
tests, between two spatially separated parts of the solid, thereby ruling out any
local classical correlations, but it is not entirely clear how to do this and we leave
this issue for future research.

\section{Conclusions}

In this paper we have been concerned with deriving an upper bound
on the amount of correlations in a solid. Our method can be linked
to one way of deriving the Bogoliubov bound. We have applied it to
two scenarios: two qubits interacting in the Heisenberg way, which
was just used to illustrate our method, and a chain of qubits
interaction in the Ising way but being placed in an external
transverse field. We have also discussed several limitations of
this method. Although the upper bound presented here can tell us a
great deal about total correlations - and therefore do also bound
entanglement, in order to gain a better understanding it would be
advantageous to find similar lower bounds. For example, the sum of
all bipartite entanglement per qubit is one such bound and it is
relatively easy to compute. In fact, for the Ising model in a
transverse field (and at zero temperature) we have the results we
need to compute this \cite{ved01,OsNi02}. It would be important,
therefore, to try to derive a tight lower bound on entanglement
(and correlations in general), and this we believe to be a very
fruitful direction for future research. Another issue is that in
general we would have to considered the spin of particles involved
\cite{ved03}. This will add the particle statistics consideration
into our picture and this, as is well known, can affect the amount
of entanglement, as well as being able to convert entanglement for
the internal degrees of freedom to the spatial (external) degrees
of freedom \cite{ved03}. Although some methods have been developed
for quantifying entanglement under those circumstances, this
relationship between ``internal" and ``external" entanglement is
still not properly understood and the problem would be well beyond
the scope of the current investigations.

{\bf Acknowledgements.} Part of the work on this paper was performed during the visit
to the Perimeter Institute, whose hospitality the author greatly acknowledges. The
author would also like to thank Julian Hartley and Jiannis Pachos for helpful
discussions and comments on this work. Julian Hartley is in addition acknowledged for
his help with plotting the figures. This work has been supported by the Engineering
and Physical Sciences Research Council, European Union and Elsag Spa.


\begin{references}

\bibitem{Nature} S. Ghosh, T. F. Rosenbaum, G. Aeppli, S. N. Coppersmith, Nature {\bf 425}, 48
(2003); V. Vedral, Nature {\bf 425}, 28 (2003).
%
\bibitem{Vedral} V. Vedral, Rev. Mod. Phys. {\bf 74}, 197 (2002).
%
\bibitem{Fumiaki} F. Morikoshi, M. Santos and V. Vedral, {\em quant-ph/0306032} (2003).
%
\bibitem{NiTe} M. A. Nielsen, (PhD Thesis, University of New Mexico, New Mexico, USA, 1998), also
{\em quant-ph/0011036}.
%
\bibitem{W00} W. K. Wootters, {\em quant-ph/0001114} (2000).
%
\bibitem{OW01} K. M. O'Connor and W. K. Wootters, Phys. Rev. A {\bf 63}, 052302 (2001).
%
\bibitem{ved01} M. C. Arnesen, S. Bose and V. Vedral, Phys. Rev. Lett. {\bf 87}, 017901 (2001).
%
\bibitem{ved02} D. Gunlycke, S. Bose, V.M. Kendon and V. Vedral,
Phys. Rev. A {\bf 64}, 042302 (2001).
%
\bibitem{WFS01} X. Wang, H. Fu and A. I. Solomon, J. Phys. A: Math. Gen. {\bf 34}, 11307 (2001).
%
\bibitem{WaZa02} X. Wang and P. Zanardi, Phys. Lett. A 301 (1-2), 1 (2002).
%
\bibitem{Wa02} X. Wang, Phys. Rev. A 66, 034302 (2002).
%
\bibitem{rest} P. Zanardi and X. Wang, J. Phys. A {\bf 35}, 7947 (2002); Yu Shi, {\em quant-ph/0204058}
and {\em cond-mat/0205272} (2002); J. Schliemann, {\em
quant-ph/0212114} (2002).
%
\bibitem{Os02} A. Osterloh, L. Amico, G. Falci and R. Fazio, Nature
{\bf 416}, 608 (2002).
%
\bibitem{OsNi02} T. J. Osborne and M. A. Nielsen, Phys. Rev. A {\bf
66}, 032110 (2002).
%
\bibitem{vid02} G. Vidal, J. I. Latorre, E. Rico and A. Kitaev,
\emph{quant-ph/0211074} (2002).
%
\bibitem{Kubo} M. Toda, R. Kubo and N. Saito, {\em Statistical Physics I} (Springer, Berlin, 1978).
%
\bibitem{Ann} E. Lieb, T. Schultz and D. Mattis, Annals of Phys. {\bf
16} 407 (1961).
%
\bibitem{Kat} S. Katsura, Phys. Rev. {\bf 127} 1508 (1962)
%
\bibitem{sach99} S. Sachdev, {\em Quantum Phase Transitions} (Cambridge
Univ. Press, 1999).
%
\bibitem{Henderson} L. Henderson and V. Vedral, J. Phys. A {\bf 34}, 6899 (2001).
%
\bibitem{Onsager} L. Onsager, Phys. Rev. {\bf 65}, 117 (1944).
%
\bibitem{Feynman} R. P. Feynman, {\em Statistical Physics} (Addison-Wesley, New York, 1972).
%
\bibitem{ved03} V. Vedral, Cent. Eu. J. Phys. {\bf 2} 289 (2003).

\end{references}
\end{document}